\documentclass[showpacs,twocolumn,amsmath,amssymb]{revtex4-1}
\usepackage{graphicx}
\usepackage{dcolumn}
\usepackage{bm}
\usepackage{subfigure}

\newcommand{\anferm}{\psi^{\phantom{\dagger}}}

\begin{document}

\title{Narrowing of topological bands due to electronic orbital degrees of freedom}

\author{J\"orn W.F. Venderbos, Maria Daghofer and Jeroen van den Brink}
\affiliation{%
 Institute for Theoretical Solid State Physics, IFW Dresden, 01171 Dresden, Germany 
}%

\date{\today}


\begin{abstract}
The Fractional Quantum Hall  (FQH) effect has been predicted to
  occur in {\it absence} of magnetic fields and at high temperature in
  lattice systems that have flat bands with non-zero Chern
  number. 
We demonstrate 
that
  orbital degrees of freedom in frustrated lattice systems lead
to a  narrowing of topologically nontrivial bands. 
This robust effect does not rely on fine-tuned  
long-range
hopping parameters 
and is directly relevant to a wide class of transition metal
compounds. 
\end{abstract}


\maketitle

Investigating the repercussions of topology on the electronic states
in condensed matter systems has a long and rich history.  The Integer
Quantum Hall (IQH) effect, discovered~\cite{Klitzing80} in 1980, was
soon  understood to be a profound 
manifestation of the topological properties of the Landau levels.
The quantized Hall conductance 
was shown to 
be a topological invariant that classifies the ground state~\cite{TKNN}.  Later that decade,
Haldane~\cite{haldane88} showed that the IQH state is not 
restricted to two-dimensional (2D) electron gases in a strong magnetic field.
It can also be realized in lattice systems without
Landau levels, by introducing electrons on a lattice with complex
hoppings that 
break time-reversal symmetry.  
In recent years, topologically nontrivial electronic phases  were
moreover discovered 
in time-reversal invariant
insulators
~\cite{PhysRevLett.95.146802,PhysRevLett.98.106803,PhysRevB.79.195322,PhysRevB.75.121306},
leading to the Quantum Spin Hall effect in 
2D~\cite{Bernevig15122006,Konig02112007}
 and to the existence of protected 2D Dirac fermions on the surface of 3D topological
insulators~\cite{RMP_Hasan_Kane,RMP_Qi_Zhang} and a related quantum
Hall effect~\cite{PhysRevLett.106.126803}.

These presently much studied  2D and 3D topological insulators are
time-reversal invariant lattice systems that can therefore be
perceived as a further generalization of the Quantum Hall states. The
generalization of Fractional Quantum Hall (FQH) states, the fractional
counterpart of the IQH states, was considered only very
recently~\cite{tang10,sun10,neupert10}.  
Such a lattice FQH effect will be quite different from the ordinary
FQH in 2D electron gases, for instance requiring a variational
wavefunction distinct from the Laughlin wave
function~\cite{Laughlin:1983,Qi11},  and it can occur without magnetic
field and potentially at high
temperature~\cite{tang10,sun10,neupert10}. 
Its realization in a material system would offer an exciting
prospect for quantum computation, since the presence of non-Abelian
FQH states allows for the creation of topologically protected
qbits~\cite{DasSarma05}. Creating the analogue of the FQH effect in a lattice
system requires the fractional filling of topologically nontrivial
bands, which should be very narrow~\cite{tang10,sun10,neupert10}, so
that the electron-electron interactions can dominate over the kinetic
energy and induce FQH states~\cite{sheng11}.

The theoretical approach used to progress toward this
goal so far relies on the fine-tuning of the electron kinetic energy 
in model Hamiltonians containing bands with the
correct topological properties~\cite{tang10,sun10,neupert10,Hu2011}. 
Such a flattening procedure of the bands 
usually requires
tuning (very) long-range hopping parameters to a set of quite peculiar
strengths, which in real materials represents a rather formidable
challenge from an experimental point of view. 

We consider orbital degrees of freedom as an alternative agent for
band flattening. Orbitals naturally occur in many transition metal
(TM) compounds, which 
at the same time  feature  
strong electron-electron
interactions~\cite{KK1982,Tokura00}. 
We concentrate on manganites, where Mn$^{3+}$ ions are in a high-spin
$3d^4$ configuration,  with three electrons in the more localized
$t_{2g}$ states forming a spin of 3/2 and one electron in either of the two more
itinerant $e_g$ orbitals ferromagnetically coupled to this spin.
Apart from other $3d$ systems besides Mn, the versatile class of TM oxides
also contains $4d$ and $5d$ materials with orbital degrees of freedom, 
of which ruthenates~\cite{1367-2630-7-1-066} and
iridates~\cite{Jackeli:2009p2016,Pesin:2010p2560} are important
examples.

We will show that in the presence of a chiral spin texture, such an
orbital make-up leads to nearly flat topologically nontrivial bands.
It is well established that geometric frustration 
may
stabilize non-coplanar spin-chiral magnetic textures when itinerant
electrons couple to localized spins~\cite{Shindou01,martin08,Akagi:2010p083711,Kumar:2010p216405,Chern10}.
The Berry phase acquired by the electrons then
leads to topologically nontrivial
bands~\cite{Ohgushi:00,Taguchi:2001p2556,Chen:2010p2591,martin08,Akagi:2010p083711}.
The pronounced spatial anisotropy of the $e_g$ and  $t_{2g}$ orbitals strongly affects the
symmetry of hopping integrals, even suppressing hopping completely
along some directions~\cite{KK1982}. This can result in very flat
bands like the dispersionless bands found in several multi-orbital TM compounds
-- a number of antiferromagnetic phases in cubic manganites are 
stabilized by such a mechanism~\cite{vandenBrink:1999p2416,Hotta:Prl}. 
Here, we report a strong orbital-induced flattening of topological bands in
spin-chiral phases on frustrated kagome and triangular lattices,
demonstrating that orbital degrees of freedom 
of transition-metal ions 
generically provide a route to realizing a lattice version of the FQH
effect. 
We also present indications that residual interactions can then
induce a FQH state.

\begin{figure}
\includegraphics[width=0.9\columnwidth]{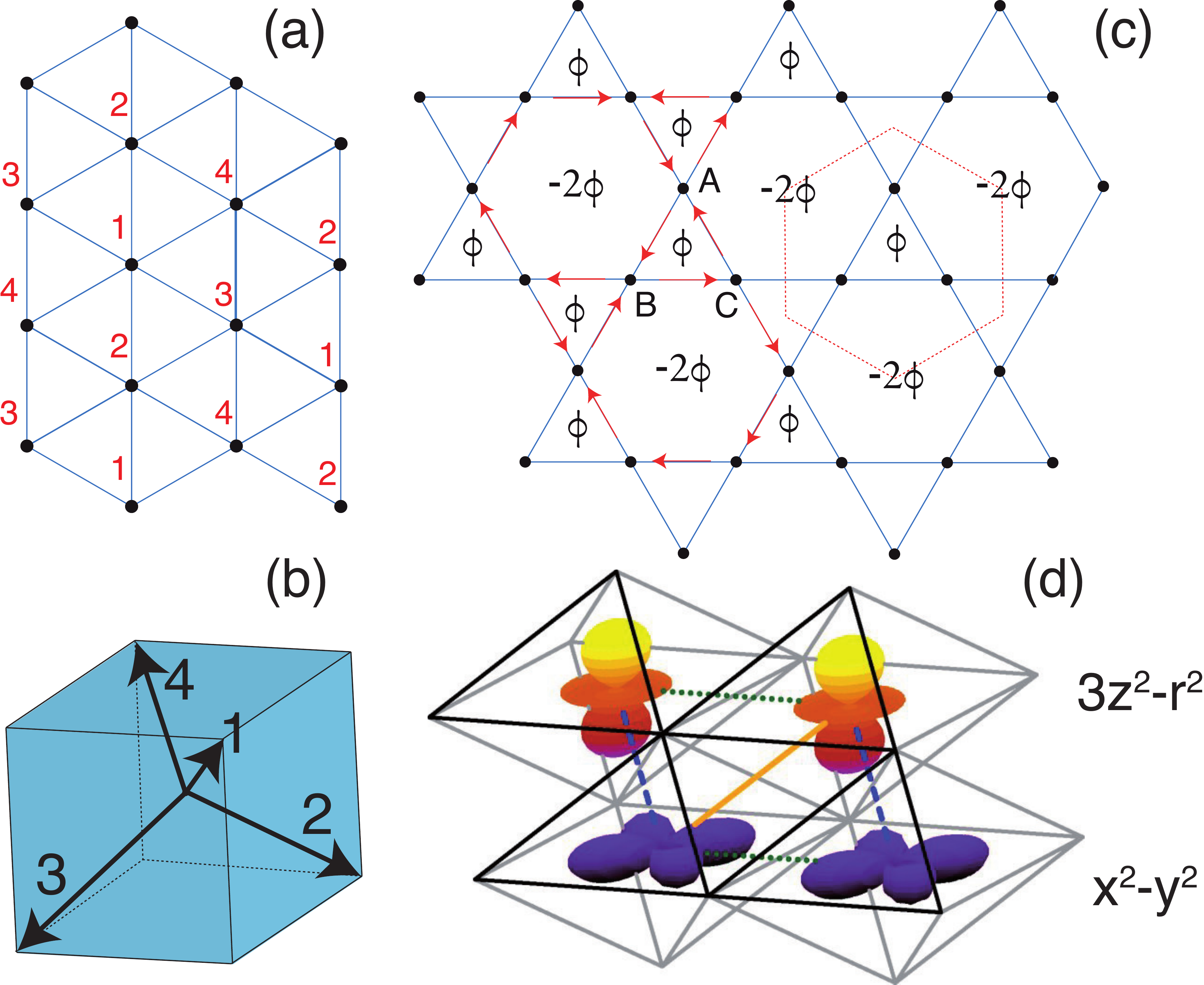}
\caption{\label{fig:lattices} (Color online) (a) Chiral spin ordering on the
  triangular lattice. 
  (b) Spins forming a regular tetrahedron, numbers refer to sites in
  (a). 
  (c) Flux-phase state on the kagome lattice. The unit cell
  is indicated by the dashed hexagon and the gauge choice by arrows;
  a flux $\phi $ threads each triangle.  
  (d) Nearest-neighbor hopping geometry in lattices with triangular
  symmetry. Grey lines illustrate 
  the oxygen octahedra, black front facets illustrate the triangular
  geometry. Thick dotted, dashed and
  solid lines indicate the bonds corresponding to the hopping matrices
  $\hat{T}_{1}$,
  $\hat{T}_{2}$, and $\hat{T}_{3}$. Two $d_{3z^2-r^2}$ (top) and
  $d_{x^2-y^2}$ (bottom) orbitals are also shown. \label{fig:orbs}}
\end{figure}

{\it Chiral spin textures---}
We first summarize the situation on the triangular and kagome lattices 
for mobile charge carriers without orbital degrees of freedom in
presence of a nontrivial spin-texture. 
The Kondo Lattice Model, which describes the interaction between
localized ($t_{2g}$) spins and itinerant ($e_g$) electrons, exhibits a
topologically nontrivial chiral spin state on the 
triangular
lattice~\cite{martin08,Akagi:2010p083711,Kumar:2010p216405}. This
state has a four--sub-lattice ordering, illustrated in
Fig.~\ref{fig:lattices}(a) and a finite scalar spin
chirality $\langle {\bf S}_1 \cdot  {\bf S}_2 \times  {\bf S}_3
\rangle \neq 0$ for spins on the corners of triangles.  
The situation then becomes equivalent to electrons hopping on
a triangular lattice with a fictitious gauge flux of $\phi = \pi/2 $
threading each triangle, see Fig.~\ref{fig:lattices}. 
The effective electronic Hamiltonian has a two-site unit cell~\cite{martin08} and
two bands, with Chern numbers $\pm 1$, separated by a gap.
On the kagome lattice, 
a staggered flux pattern, shown in Fig.~\ref{fig:lattices}, can result
from topologically nontrivial spin states,   
where a flux $\phi$
threads each triangle and a flux $-2\phi$ each hexagon~\cite{Ohgushi:00}.
Time-reversal symmetry is broken for
$\phi\neq 0,\pi$ and two gaps open, leading to three bands. 
The middle band has zero Chern number, but the lowest and highest
are topologically nontrivial with $C=\mp\text{sgn}(\sin \phi)$~\cite{Ohgushi:00}.  
However, all topologically nontrivial bands  
have a considerable dispersion
on both lattices, 
which we will now show to be substantially reduced
by the presence of an orbital degree of freedom.

\begin{figure}
\includegraphics[width=\columnwidth]{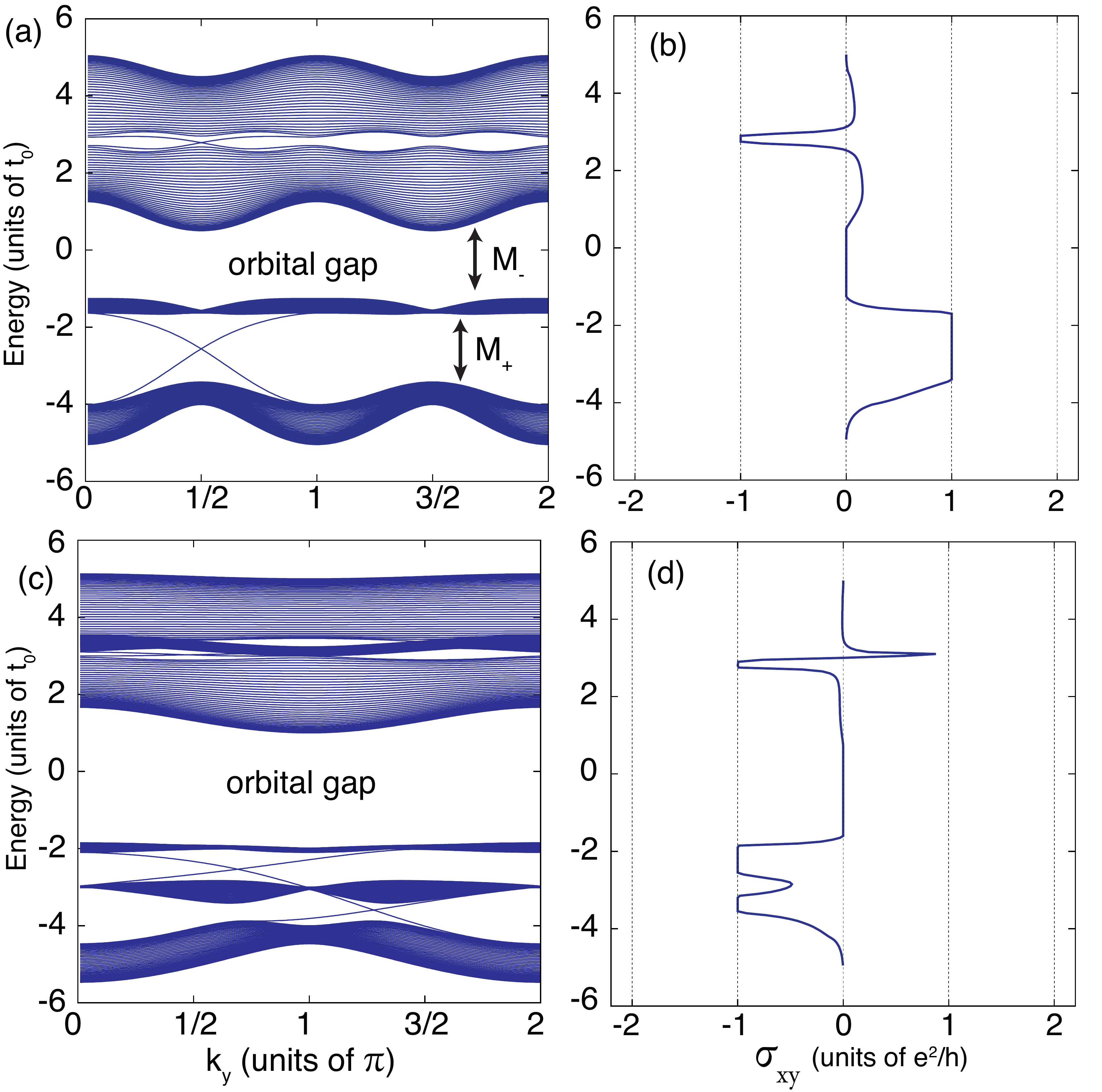} 
\caption{\label{fig:bands_sigma} Top: kagome lattice, bottom: triangular lattice. Left: bands of a strip geometry, clearly showing the chiral edge states of the flattened bands. Right: off-diagonal Hall conductivity as function of chemical potential, providing the Chern numbers for the flat bands. 
For the Kagome lattice $t' = -0.46$, $\Delta=2.75$, $\phi=\pi/4$ and for the triangular case $t' = -0.46$, $\Delta=2.5$.}
\end{figure}

{\it Orbital degree of freedom---}
In many TM oxides, the TM ions are inside oxygen octahedra, which are
edge-sharing in a triangular lattice, 
see Fig.~\ref{fig:orbs}. The cubic symmetry
splits the TM $d$ levels into three $t_{2g}$ and two $e_g$ orbitals.
We focus mainly on the latter, but later also demonstrate an analogous
effect for the former.
Along the ${\bf a_1}=(0,1)$ direction, indicated by a dotted line in
Fig.~\ref{fig:orbs}, hopping for $e_g$ orbitals conserves orbital
flavor and is given  
by $t$ ($t'$) for the $|x^2-y^2 \rangle$ ($|3z^2-r^2 \rangle$)
orbital. Hoppings along the other two bonds are obtained by a rotation
in orbital space, the hopping matrices along all three bonds are given
by Eq.~(1) in~\cite{suppl}.
We use $t$ as unit of energy and vary the material-dependent
ratio $t'/t$ between $-1$ and 1, concentrating on $t'/t<0$ inferred
from direct overlaps of the orbitals~\cite{note_hopp_sk}. 
For our fillings of approximately one electron per site, the
Jahn-Teller effect is important and can 
induce a uniform crystal field  $H_{\textrm{JT}} = \Delta (n_x-n_z) $,
lifting orbital degeneracy~\cite{note_hopp_JT}.

The two sites of the unit cell in the spin-chiral
phase~\cite{martin08} together with the two $e_g$ orbitals give a
$4\times 4$ Hamiltonian in momentum space, see~\cite{suppl}.
Figure~\ref{fig:bands_sigma} shows the energy bands calculated for a
strip geometry, with 
periodic boundary conditions in one direction and two edges in the
orthogonal direction. 
For a large enough crystal-field splitting $\Delta> t, t'$, a gap
separates bands with different orbital character. Within each
subsystem, chiral magnetic order induces a further splitting into
two topologically nontrivial bands with Chern numbers $C=\pm 1$. This
is unambiguously indicated by the topological edge states connecting
the bands with $C>0$ and $C<0$ in Fig.~\ref{fig:bands_sigma}(a). The transverse Hall conductivity
$\sigma^n_{xy}$ is shown in
Fig.~\ref{fig:bands_sigma}(b) and directly reflects
the topological character of the bands.

\begin{figure}
\includegraphics[width=\columnwidth]{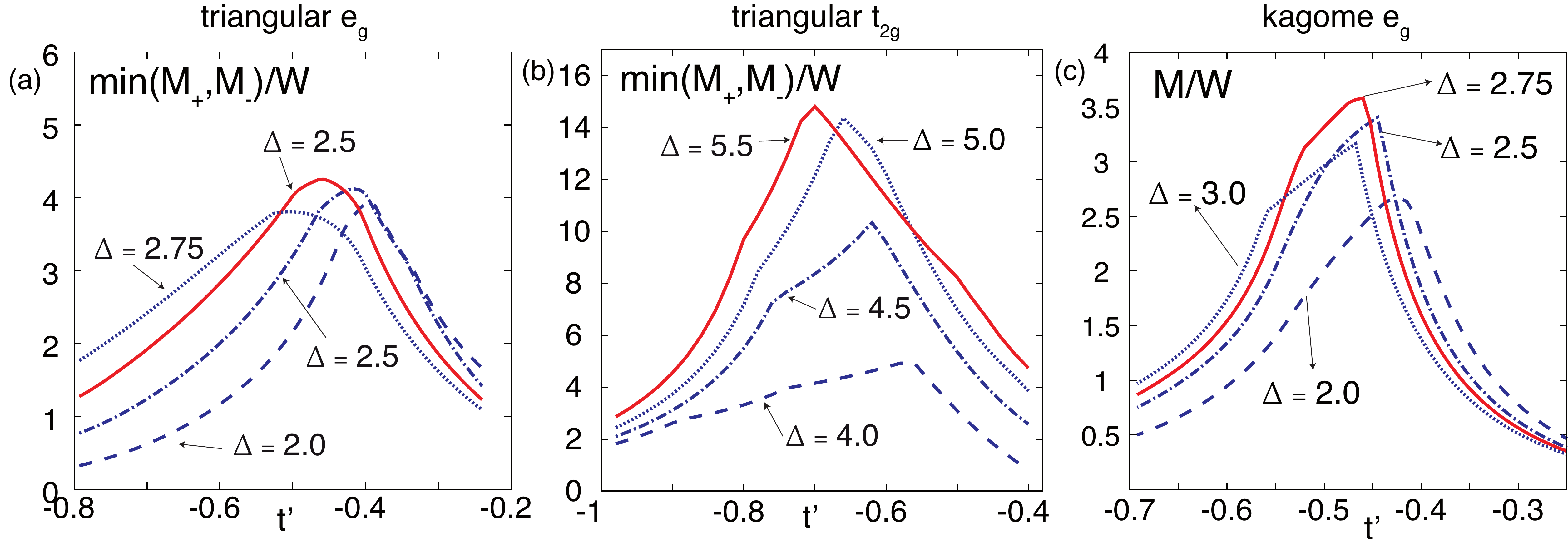}
\caption{\label{fig:gaps_figs} (Color online) (a)  The smallest band gap over bandwidth ratios for $e_g$ electrons 
  the triangular lattice, for the gaps $M_+$ and $M_{-}$ indicated in
  Fig.~\ref{fig:bands_sigma}(a). (b) The same for $t_{2g}$ electrons on a triangular lattice and (c) for $e_g$ on the kagome lattice. Orbital splittings are indicated
  by $\Delta$.}
\end{figure}

The gap between the topological bands is smaller in the
upper ( $x^2-y^2$) sector, but robust
between the two $3z^2-r^2$ bands below the crystal-field gap.  
The upper band of the $3z^2-r^2$ sector with $C=-1$, has a weak dispersion,
becoming nearly flat for $t'\approx -t/2$, see Fig.~\ref{fig:bands_sigma}(a).
The figure of merit quantifying the flatness is the ratio of the gaps $M$
separating it from other bands to its band width $W$.
Here we monitor both the `topological' gap ($M_+$), which is induced by
chiral order and the `trivial' crystal-field gap ($M_{-1}$), which separates
it from the
$x^2-y^2$ sector above. 
Figure~\ref{fig:gaps_figs}(a) shows these ratios depending on $t'$ and
$\Delta$, the relevant figure of 
merit is the smaller of the two ratios $M_+/W$ and $M_{-1}/W$. It is
appreciable in a broad range of 
$\Delta$ and $t'$, reaching a maximum of $\sim4.25$ for
$\Delta = 2.5$ and $t' = -0.45$.  

We can also consider $t_{2g}$ orbitals $|xy \rangle$, $|xz \rangle$,
and  $|yz \rangle$. The hopping matrices are given in Eq.~(2) of~\cite{suppl} and
consist of inter-orbital hopping $t$ (primarily via
ligand oxygen ions~\cite{Pen97}) and orbital-conserving hopping $t'$ due
to direct overlap~\cite{Koshibae03}. In a three-fold symmetry, the
$t_{2g}$ manifold is further split into one $a_{1g}$ and two $e_{g}^{\prime}$
states separated by a crystal field $H_{\textrm{JT}}$~\cite{JT_tri}. 
Qualitatively, we find similar behavior as for $e_g$ orbitals, but the
figure of merit reaches $M/W \approx 14$, see 
Fig.~\ref{fig:gaps_figs}(b). This extraordinarily  
large $M/W$ reflects the very large spatial anisotropy 
of $t_{2g}$ hopping integrals, well-known in 
triangular
vanadates~\cite{Pen97} and cobaltates~\cite{Koshibae03}.  

After discussing the triangular lattice, we now come to $e_g$ orbitals
on the kagome lattice illustrated in Fig.~\ref{fig:lattices}(b). 
The lattice has a three-site unit cell, and one proceeds with an
approach analogous to the two-site unite cell discussed above, for
details see~\cite{suppl}.
Bands are again separated by
the crystal field into two parts with a different orbital character,
see Fig.~\ref{fig:bands_sigma}.    
While the $x^2-y^2$ sector is hardly gapped, the $3z^2-r^3$ sector
shows three sub-bands with $C=0,\pm1$ similar to the one-band
model~\cite{Ohgushi:00}. Again, the topological character can be inferred from
edge states and is confirmed by the Hall conductivity shown in
Fig.~\ref{fig:bands_sigma}(d) and as for the triangular lattice, the top band of the
$3z^2-r^3$ sector (with $C=+1$) is very
flat for $t'\approx -t/2$. 
The figure of merit  $M/W$ is plotted in Fig.~\ref{fig:gaps_figs}(c);
it reaches values up to 
$\sim 3.5$ for $\Delta =2.75$ and $t'=-0.45$, compared to
$W/M\lesssim 1$ in the one-band model without orbital degrees of
freedom.

\begin{figure}
\includegraphics[width=\columnwidth]{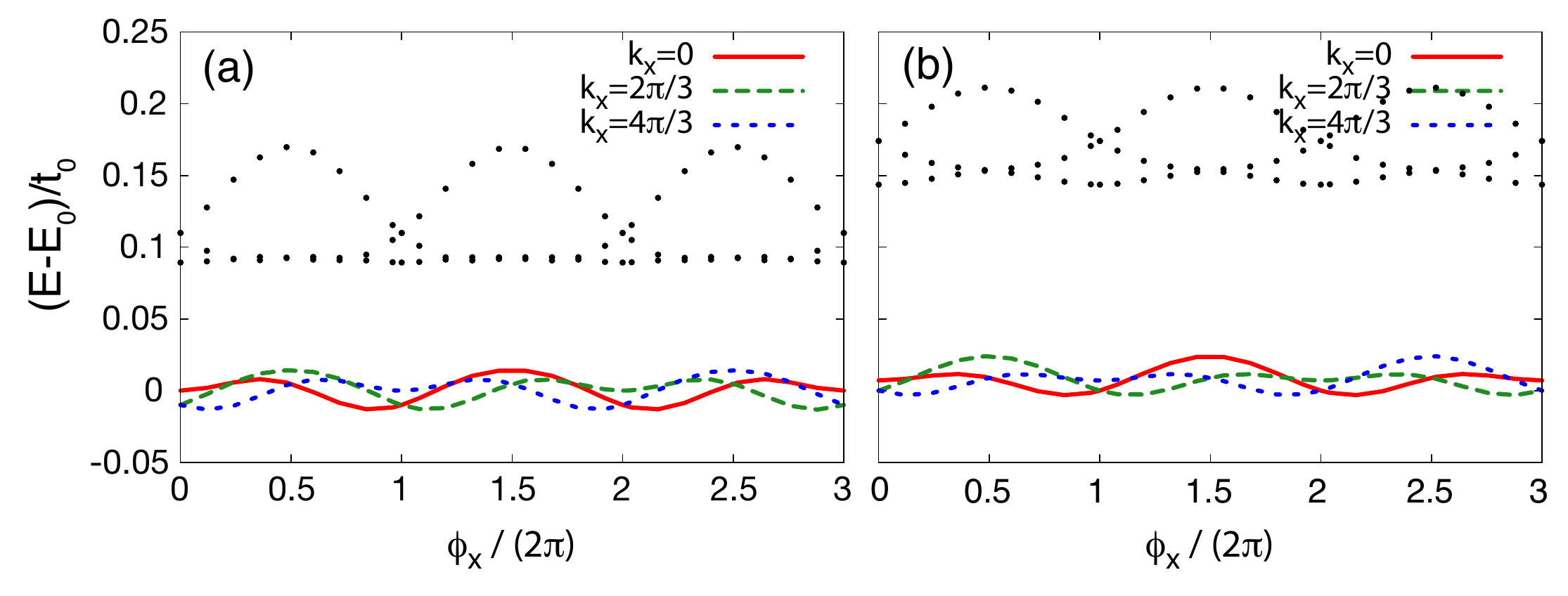} 
\caption{\label{fig:ED_panel} (Color online) Lowest energies of the
  interacting $e_g$ system on the triangular lattice, depending on flux inserted for (a) $U=2$, $V=1$,
  and (b) $U=4$, $V=2$. The ED results presented here were
  obtained for a $2\times 6 $ system with $t'/t=-0.46$ and $\Delta = 1.5$, see~\cite{suppl}.} 
\end{figure}

{\it FQH state induced by residual interactions---} 
After showing that orbital degrees of freedom can lead to flat bands
with Chern number $C\neq 0$, we analyze the impact of Coulomb
repulsion on the triangular-lattice $e_g$ system. 
We include electron-electron interactions 
\begin{equation}
H_{\text{int}} = U\sum_{i} n_{i,x^2-y^2}n_{i,3z^2-r^2} + V\sum_{\langle i,j \rangle, \alpha\beta} n_{i,\alpha}n_{j,\beta}\;,
\end{equation}
where $n_{i,\alpha}$ is the electron density in orbital $\alpha$ on
site $i$, $U$ acts on electrons occupying two orbitals on the same
site, and $V$ gives the nearest-neighbor (NN) interaction. Following
Refs.~\cite{neupert10,sheng11}, we use Exact Diagonalization
(ED) to study signatures of FQH-like states for a filling $1/3$ of the
topologically nontrivial flat band. 
For details on the parameters and the method, see~\cite{suppl}.
We find an approximate threefold ground-state manifold,
which is an indication for the topological degeneracy of a FQH
state. The three ``ground states'' cross each other upon inserting a 
magnetic flux\cite{thouless89}, see  Fig.~\ref{fig:ED_panel} and \cite{suppl}, in
agreement with results for other models~\cite{neupert10,regnault11, Wang2011}.

{\it Discussion---}
The flattening due to the orbital degrees of freedom presented here can 
be further enhanced by introducing and fine-tuning longer-range hopping integrals,
as in one-band models. Perfectly flat and topologically
nontrivial bands  ($M/W \rightarrow
\infty$) can in principle be obtained by allowing for arbitrarily
long-range hoppings~\cite{sun10}.  
We do not explore this here, 
as we aim to show that the anisotropy inherent in $d$-orbital degrees of
freedom can robustly flatten topological bands even with purely NN 
hopping, and obtain flattening ratios up to $M/W\approx 4$
($M/W\approx 14$) for
$e_g$ ($t_{2g}$) systems. 

Orbital degrees of freedom are directly relevant to
numerous well-known TM systems. Manganese compounds alone, 
 which closely correspond to the $e_g$ model
 studied here, 
occur in a variety of crystal structures: in simple
cubic or square lattices in La$_{1-x}$Sr$_x$MnO$_3$ and LaSrMnO$_4$,
honeycomb in e.g. Li$_2$MnO$_3$ or (Bi$_3$Mn$_4$O$_{12}$)NO$_3$ but also in
strongly frustrated pyrochlore lattices as in
e.g. Ti$_2$Mn$_2$O$_7$ and in triangular lattices as in YMnO$_3$. The pyrochlore lattice, in particular, is
realized by the $B$-site TM ions in the very common spinel crystal
structure, and can be
thought of as consisting of kagome and triangular layers stacked along
the $(1,1,1)$ direction.  
Singling out the triangular or kagome layers,
possibly via chemical substitution or controlled monolayer growth, 
may thus
lead to systems similar to the ones studied here. 

A FQH ground
state needs an interaction $V$ that exceeds the bandwidth ($V>W$), but remains smaller
than the gap ($V<M$) so that bands are not mixed. 
Fortunately, TM oxides have substantial
Coulomb interactions and NN repulsion $V$
can become as large as or larger than the hoppings. On the
other hand, it is almost always smaller than onsite repulsion, which 
increases one of the two gaps delimiting the flat
band~\cite{suppl}. The remaining challenge is thus to keep the
effective interaction 
small compared to the gap separating the bands with $C=\pm
1$. A large figure of merit $M/W$ provides a large window for this
separation of energy scales $W<V<M$ and our ED results in Fig.~\ref{fig:ED_panel}
suggest that one indeed has some flexibility in this
regard, as interactions differing by a factor of two lead to similar
results. The crystal field splitting needed to obtain the desired
flattening can be  varied by applying (chemical) pressure. 
Finally, many of these materials, manganites in particular, are
easy to dope which allows control of the (fractional) band
filling.  

Theoretically, there is no fundamental objection to the realization of
lattice FQH states, but it remains an intriguing challenge from an 
experimental and practical point of view.
We have demonstrated here that $d$-orbital
degrees of freedom, ubiquitous in TM  
compounds, substantially narrow the topologically nontrivial bands of
electrons moving in a background of non-coplanar spins. The separation
of energy scales is comparable to that achievable by
long-range hopping, and we find signatures of a FQH-like
ground state. 
In the search for the lattice FQH effect, geometrically frustrated TM compounds with an orbital degree of
freedom thus come to the fore as a promising class of candidate systems.  

{\it Acknowledgements---} This research was supported by the
Interphase Program of the Dutch Science Foundation NWO/FOM (J.V. and
JvdB) and the Emmy Noether program of the DFG (M.D.) in Germany.

\appendix

\section{Hamiltonians} \label{sec:ekin}
{\it The appendices presented here correspond to the supplementary material of the published version.} 
\subsection{Hoppings and Crystal fields in orbital space}

For $e_g$ orbitals, the hopping matrices written 
in the $( |3z^2-r^2 \rangle,  |x^2-y^2 \rangle )^T = ( |z \rangle,
|x \rangle )^T $ basis are 
\begin{gather} 
 \hat{T}_{1} =  \begin{pmatrix}
t' & 0 \\
0& t \\
\end{pmatrix}, \qquad \hat{T}_{2} = \frac{1}{4} \begin{pmatrix}
t'+3t & \sqrt{3}(t'-t) \\
 \sqrt{3}(t'-t) & 3t'+t \\
\end{pmatrix}, \nonumber\\ 
\hat{T}_{3} =  \frac{1}{4} \begin{pmatrix}
t'+3t & -\sqrt{3}(t'-t) \\
 -\sqrt{3}(t'-t) & 3t'+t \\
\end{pmatrix},\label{eq:hop}
\end{gather}
along ${\bf a_1}$, ${\bf a_2}$, and ${\bf a_3}$, respectively. 

For the $t_{2g}$ orbitals with the  basis $( |xy \rangle,  |xz \rangle,
|yz \rangle ) $, the hopping matrices are 
\begin{gather} \label{eq:hopp_t2g}
 \hat{T}_{1} =  \begin{pmatrix}
t' & 0 & 0\\
0  & 0 & t \\
0  & t & 0 \\
\end{pmatrix}, \quad 
\hat{T}_{2} = \begin{pmatrix}
0  & 0 & t\\
0  & t'& 0 \\
t  & 0 & 0 \\
\end{pmatrix}, \quad 
\hat{T}_{3} =  \begin{pmatrix}
0  & t & 0\\
t  & 0 & 0 \\
0  & 0 & t'\\
\end{pmatrix}
\end{gather}

The onsite term representing the crystal field splits the orbital
degeneracy; for the $e_g$ orbitals, it is simply (see main text)
\begin{align}\label{eq:JT_eg}
H_{\textrm{JT}} = \Delta (n_x-n_z) = \begin{pmatrix}
-\Delta  & 0 \\
0  & \Delta \\
\end{pmatrix}\;.
\end{align}
For the $t_{2g}$ system, the crystal field related to elongating or
shortening the octahedra is naturally expressed in the $\{a_{1g},
e_{g+}^{\prime}, e_{g-}^{\prime}\}$ basis reflecting the triangular symmetry. 
It is then given by $H_{\textrm{JT}} =  \Delta (n_{e_{g+}}+n_{e_{g-}}
-2n_{a_{1g}})/3$, which can be expressed in the original
$\{xy,xz,yz\}$ basis via the basis transformation defined in
Ref.~\onlinecite{Koshibae03} and becomes
\begin{align}\label{eq:JT_t2g}
H_{\textrm{JT}} = \frac{\Delta}{3} (n_{e_{g+}}+n_{e_{g-}} -2n_{a_{1g}}) = -\frac{\Delta}{3}\begin{pmatrix}
0& 1  & 1 \\
 1  & 0& 1 \\ 
 1  & 1& 0 \\ 
\end{pmatrix}\;.
\end{align}

\subsection{Hamiltonians on the Triangular and Kagome lattices}

Even though the spin pattern underlying the chiral spin state on the
triangular lattice has a four-site unit cell, the electronic
Hamiltonian has only a two-site unit cell and can thus be described
by a $2\times 2$ matrix if the original model is a one-band
model.~\cite{martin08} With an underlying multi-orbital Hamiltonian,
this becomes a Hamiltonian 
matrix consisting of four blocks ($H_{11}, H_{12}, H_{21}$ and $H_{22}$),  
where the blocks refer to the lattice sites of the spatial unit
cell and are matrices in orbital space, leading to 
\begin{align}\label{eq:Hchiral}
H &=  \begin{pmatrix}
H_{\textrm{JT}}+ H_{11} & H_{12} \\
H_{21} & H_{\textrm{JT}}+ H_{22}
\end{pmatrix}\qquad\textrm{with}\\ \nonumber
H_{11} &= -H_{22}= -2 \hat{T}_{1}\cos({\bf k}\cdot {\bf a_3}), \\ \nonumber
H_{12} &= H_{21}^{\dagger}=
 - 2\hat{T}_{2}\cos({\bf k}\cdot {\bf a_1}) -2i \hat{T}_{3} \cos({\bf
  k}\cdot {\bf a_2}) 
\end{align}
and  ${\bf a_1}+{\bf a_3}={\bf a_2}$. The hopping matrices $T_i$ are
the $2\times 2$ matrices Eq.~(\ref{eq:hop}) for $e_g$ electrons and
the $3\times 3$ matrices Eq.~(\ref{eq:hopp_t2g}) for the $t_{2g}$
system. The matrices $H_{\textrm{JT}}$ given by the crystal field
refer to Eqs.~(\ref{eq:JT_eg}) and~(\ref{eq:JT_t2g}), correspondingly.

The  lattice vectors of the kagome lattice are 
${\bf a_1} = (-1/2,-\sqrt{3}/2) $, ${\bf a_2} = (1,0) $ and ${\bf a_3}
= (-1/2,\sqrt{3}/2) $; they  connect the three sites $A$, $B$
and $C$ of the unit cell, see Fig. 1(c) of the main text.
With the two $e_g$ orbitals per site/momentum, the basis in momentum space
becomes $\anferm({\bf k}) = (\anferm_{Az}({\bf k}),\anferm_{Ax}({\bf
  k}),\anferm_{Bz}({\bf k}),\anferm_{Bx}({\bf k}),\anferm_{Cz}({\bf
  k}),\anferm_{Cx}({\bf k}))^T$. 
The hoppings between the octahedra forming the kagome lattice have
triangular geometry as well,
and due to the three-site unit cell in real space, the Hamiltonian 
consists of nine blocks, where each block is again given by a
matrix taken from  Eq.~(\ref{eq:hop}) within the  $e_{g}$-orbital space.
This leads to 
\begin{widetext}
\begin{align}
H &=  \begin{pmatrix}
M_0& M_1  &  M_3^\dagger\\
M_1^\dagger &M_0  & M_2  \\
M_3 &M_2^\dagger & M_0
\end{pmatrix} 
 =  \begin{pmatrix}
H_{\textrm{JT}}& 2 \hat{T}_{1}\cos({\bf k}\cdot {\bf a_1}) \textrm{e}^{-i\phi/3} &  2 \hat{T}_{3}\cos({\bf k}\cdot {\bf a_3}) \textrm{e}^{i\phi/3}\\
2 \hat{T}_{1}\cos({\bf k}\cdot {\bf a_1}) \textrm{e}^{i\phi/3} & H_{\textrm{JT}} &2 \hat{T}_{2}\cos({\bf k}\cdot {\bf a_2}) \textrm{e}^{-i\phi/3}  \\
2 \hat{T}_{3}\cos({\bf k}\cdot {\bf a_3}) \textrm{e}^{-i\phi/3} &2 \hat{T}_{2}\cos({\bf k}\cdot {\bf a_2}) \textrm{e}^{i\phi/3} & H_{\textrm{JT}}
\end{pmatrix}.
\end{align}
\end{widetext}
The diagonal blocks $M_0=H_{\textrm{JT}}$ refer to the crystal field
splitting Eq.~(\ref{eq:JT_eg}). 

\subsection{Interacting Hamiltonian}

In our Exact diagonalization (ED, see below) studies, we add to the
kinetic energy Eq.~(\ref{eq:Hchiral}) of the $e_g$ system on the
triangular lattice the Coulomb interaction terms as given
in the main text. It should
be noted that the nearest-neighbor interaction $V$ is defined on the original
underlying triangular lattice, i.e., the same interaction $V$ acts
both between the two real-space sites of the same unit cell and between
sites in different unit cells, if these sites are nearest-neighbors on
the triangular lattice. When adding these interactions to the
Hamiltonian, we have to take into account that $U$ implies an
additional energy cost for electrons in the energetically higher [see
Fig. 2(a) of the main text]
$x^2-y^2$ orbitals, thus changing the crystal field splitting
$\Delta$. In our ED calculations, we therefore reduce $\Delta$ from
the value $\Delta=2.5$ that would give the flattest band.

\section{Methods} \label{sec:methods}

\subsection{Chern number calculation} 
In the main text we present the Hall conductivity as a proof of the topological nature of the bands. The topological invariant characterizing the bands, the Chern number $C_n$ (where $n$ is a band label), is related to the transverse Hall conductivity as 
\begin{align}
\sigma^n_{xy} &= -i \sum_{\stackrel{{\bf k}}{m\neq m}} \frac{\langle m
  {\bf k} |J_x | n {\bf k}\rangle \langle n {\bf k} |J_y | m {\bf
    k}\rangle -\textrm{h.c.}}{(E_n({\bf k})-E_m({\bf k}))^2}f(E_n({\bf
  k}))\nonumber \\
&=  \frac{e^2}{h}C_n= \frac{-ie^2}{2\pi h}
\int_{BZ} d^2{\bf k}\; \hat{z} \cdot \nabla_{k}
\times {\bf A}_n({\bf k}),
\end{align}
where $J_i$ is the current operator in the $i$-direction, given by $J_i= \sum_k c^\dagger_k (\partial H(k)/ \partial k_i) c_k $ and $f$ is the Fermi-Dirac function. ${\bf A}_n({\bf k})$ is the so-called Berry connection as function of momentum ${\bf k}$ and defined by ${\bf A}_n({\bf k}) = -i \langle u_n({\bf k}) | \nabla_{\bf k}|u_n({\bf k}) \rangle $.

\subsection{ Exact diagonalization } \label{sec:ed}

Exact diagonalization studies 
were only performed for the $e_g$ triangular lattice model, because
this case has the smallest unit cell of  the models considered
here. Even so, only very small systems are accessible. We use $N_x
\times N_y$ real-space sites, giving $N_c = N_s/2 = N_x \times N_y/2$
unit cells of the chiral phase, but  $N_{sp} = 2N_s$ ``generalized'' orbitals
determining the size of the Hilbert space. There is no spin degree of
freedom, i.e., we study spinless fermions. The filling is the one 
expected to correspond to the simplest FQH state, i.e., $\nu=1/3$ filling
of the nearly flat band. Since this band is close to half filling,
this also implies a total filling of $1/3$, i.e., an electron number
of $N_e = N_{sp}/3$. 

Due
to the large number of ``generalized'' orbitals per unit cell and due
to the fact that our total electron filling is closer to
half filling (which additionally increases the size of the Hilbert
space), the lattice sizes accessible are considerably smaller 
than for models considered previously.~\cite{neupert10,sheng11} Since
we need the total number of sites to be divisible by 3 in order to
obtain $1/3$ filling and since we also want bot $N_x$ and $N_y$ to be
even, we consider a $2\times 6$ ``ladder'' ($N_c = 6$, $N_{sp}=24$). For
simplicity, we ``distort'' the triangular lattice and assume the $x$
direction to lie along ${\bf a_1}$ and the $y$ direction along ${\bf
  a_2}$ to be orthogonal, the third nearest-neighbor bond is still determined by ${\bf a_3}
= {\bf a_1}-{\bf a_2} $.

\begin{figure}
\includegraphics[width=0.95\columnwidth]{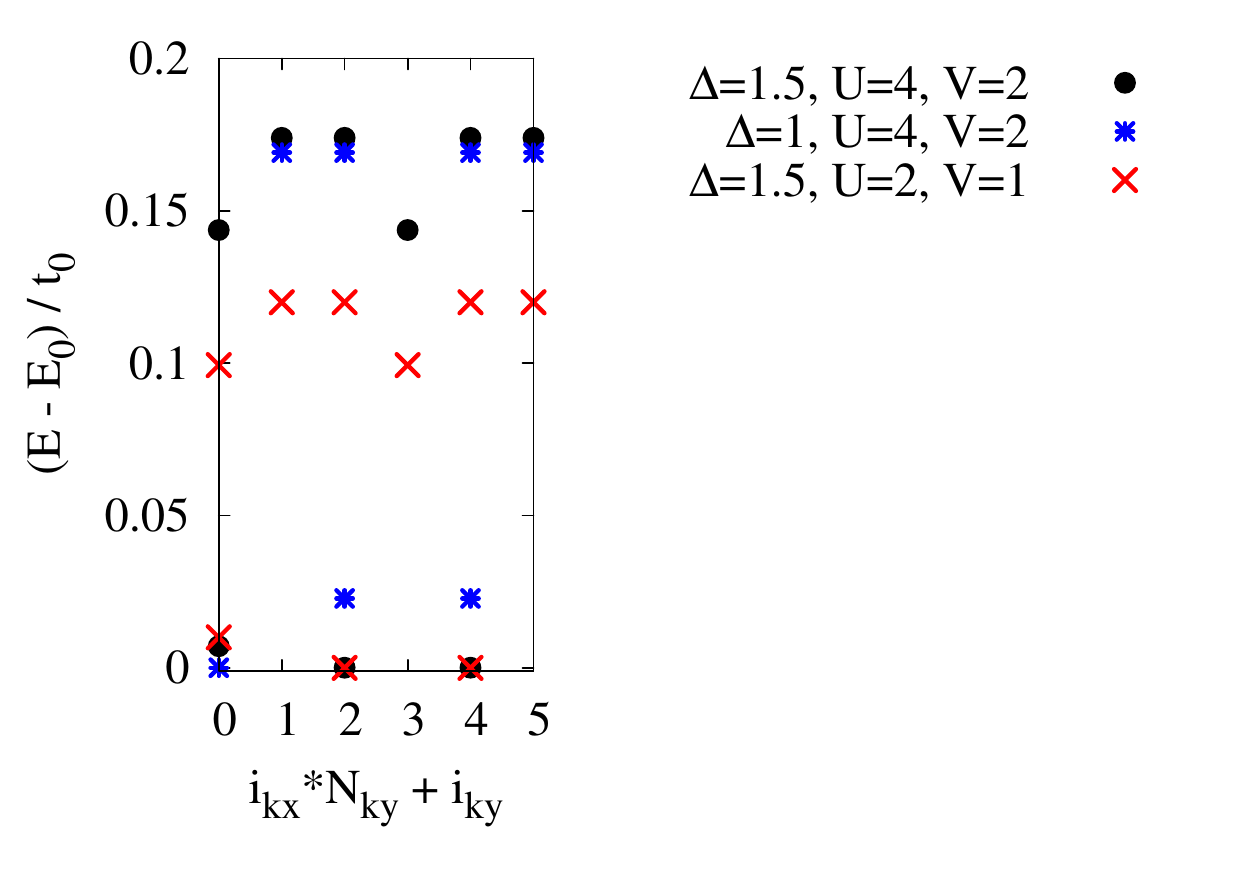}
\caption{\label{fig:ed_k} (Color online) Lowest energies of the
  interacting $e_g$ system on the triangular lattice for the various
  total momenta of the $6\times 2$ lattice for a few parameter sets, $t'/t=-0.46$} 
\end{figure}

\begin{figure}
\subfigure[$\;\Delta=0.5$, $U=4$,$V=2$]
{\includegraphics[width=0.47\columnwidth]
{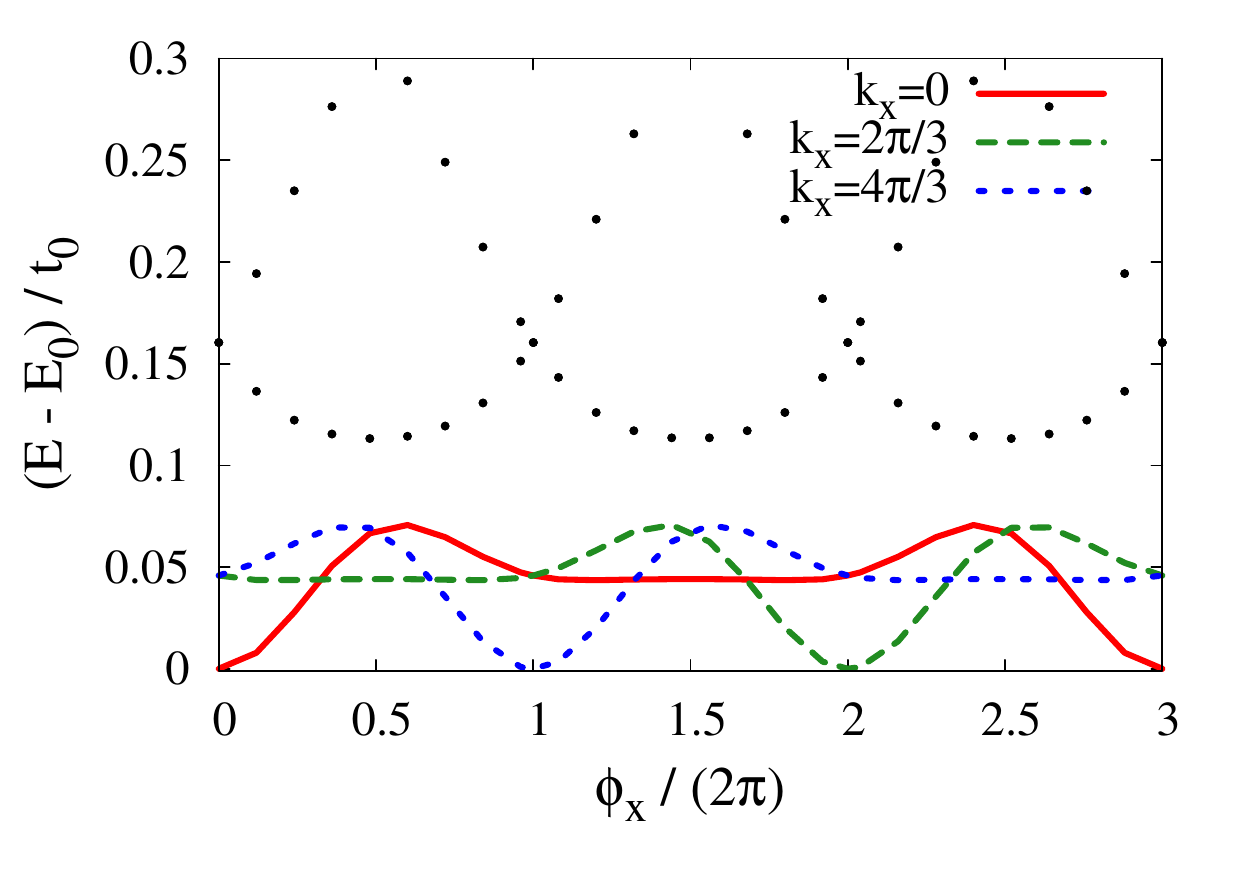}\label{fig:D05_U4_V2}}
\subfigure[$\;\Delta=1$, $U=4$, $V=2$]
{\includegraphics[width=0.47\columnwidth]
{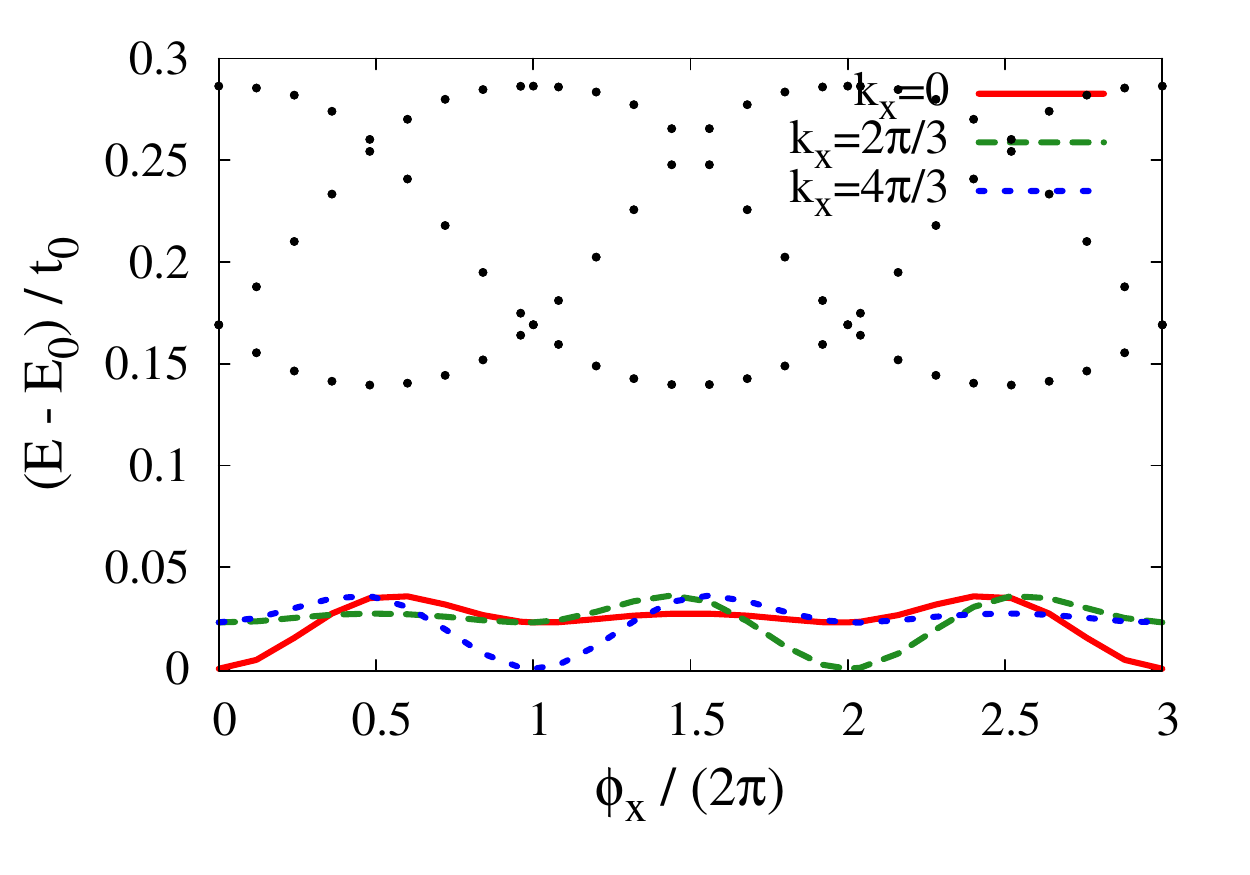}\label{fig:D1_U4_V2}}\\
\subfigure[$\;\Delta=1.5$, $U=4$, $V=2$]
{\includegraphics[width=0.47\columnwidth]
{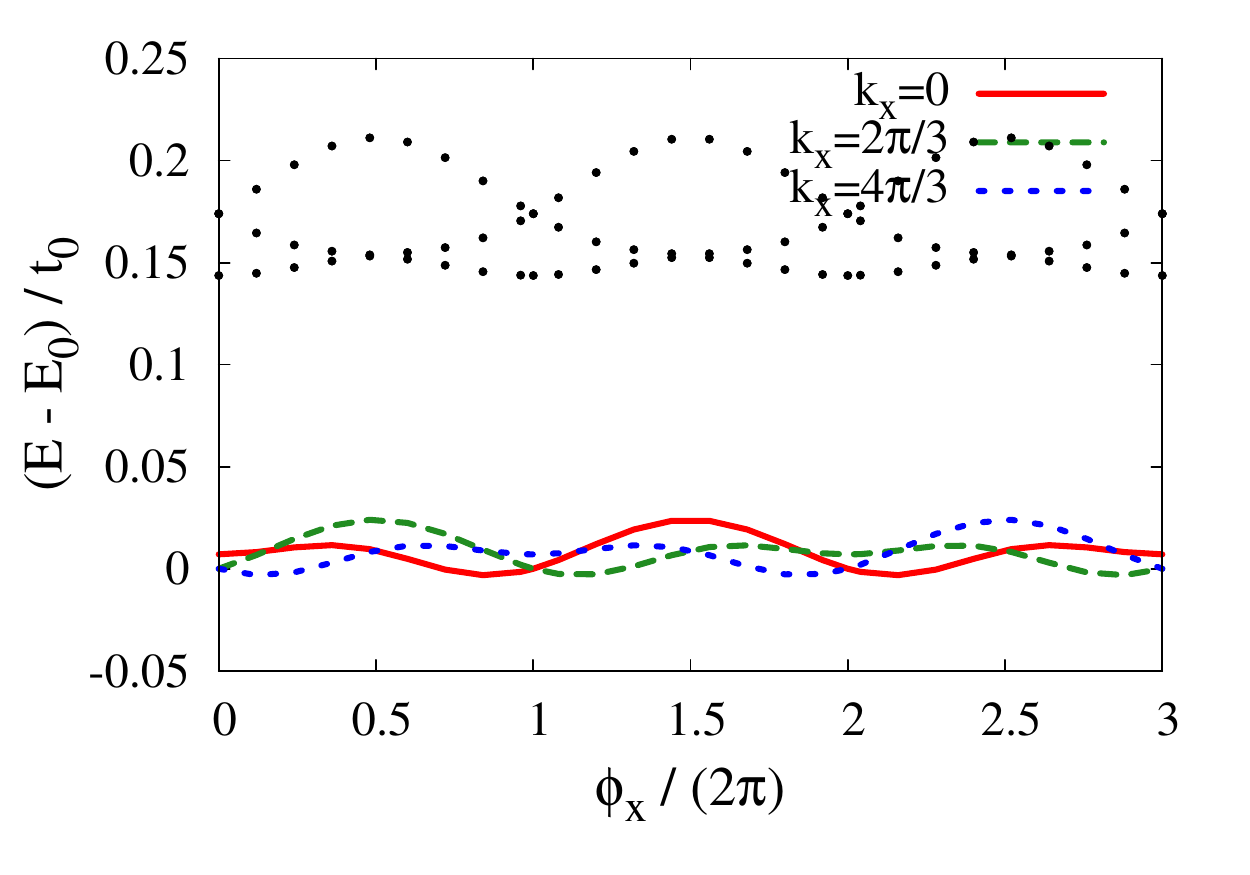}\label{fig:D15_U4_V2}}
\subfigure[$\;\Delta=2$, $U=4$, $V=2$]
{\includegraphics[width=0.47\columnwidth]
{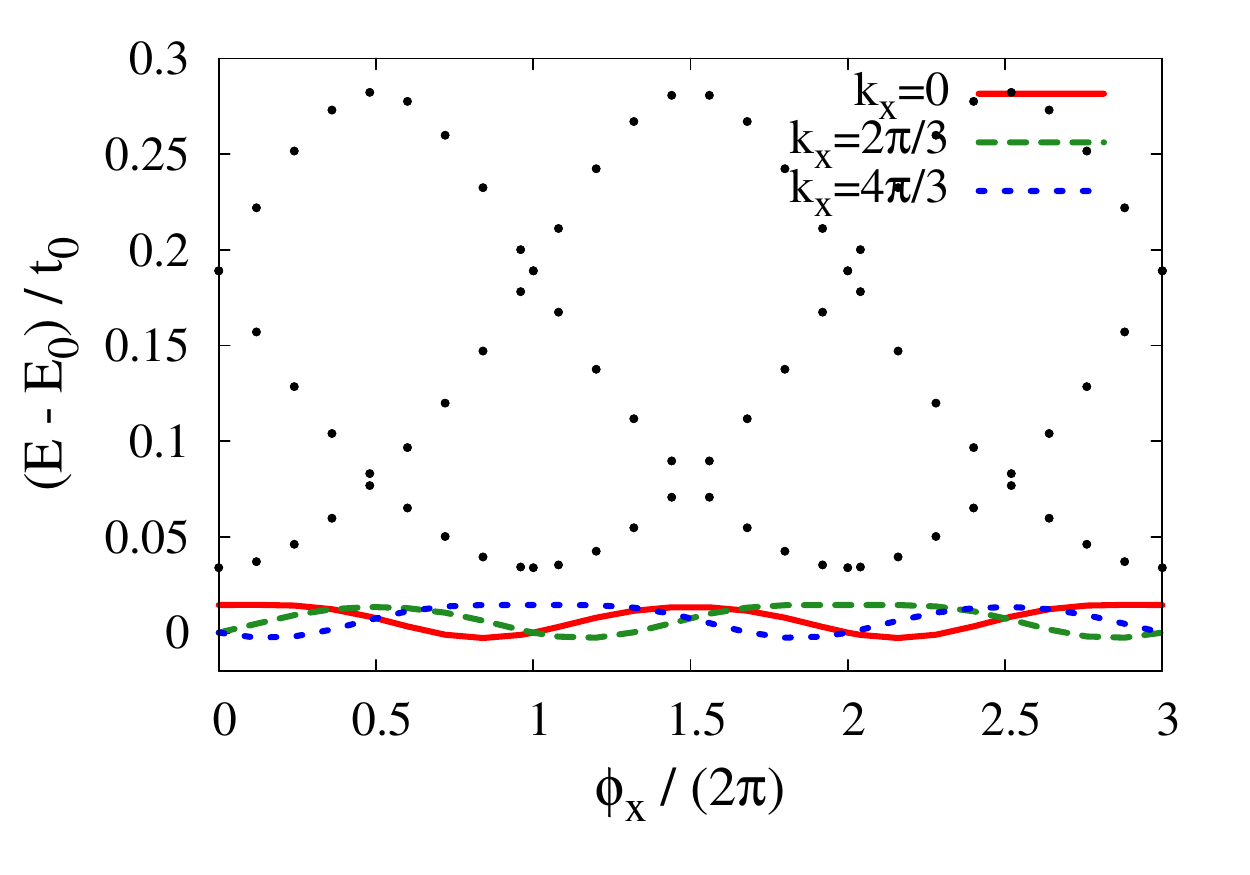}\label{fig:D2_U4_V2}}\\
\subfigure[$\;\Delta=1.5$, $U=2$, $V=1$]
{\includegraphics[width=0.47\columnwidth]
{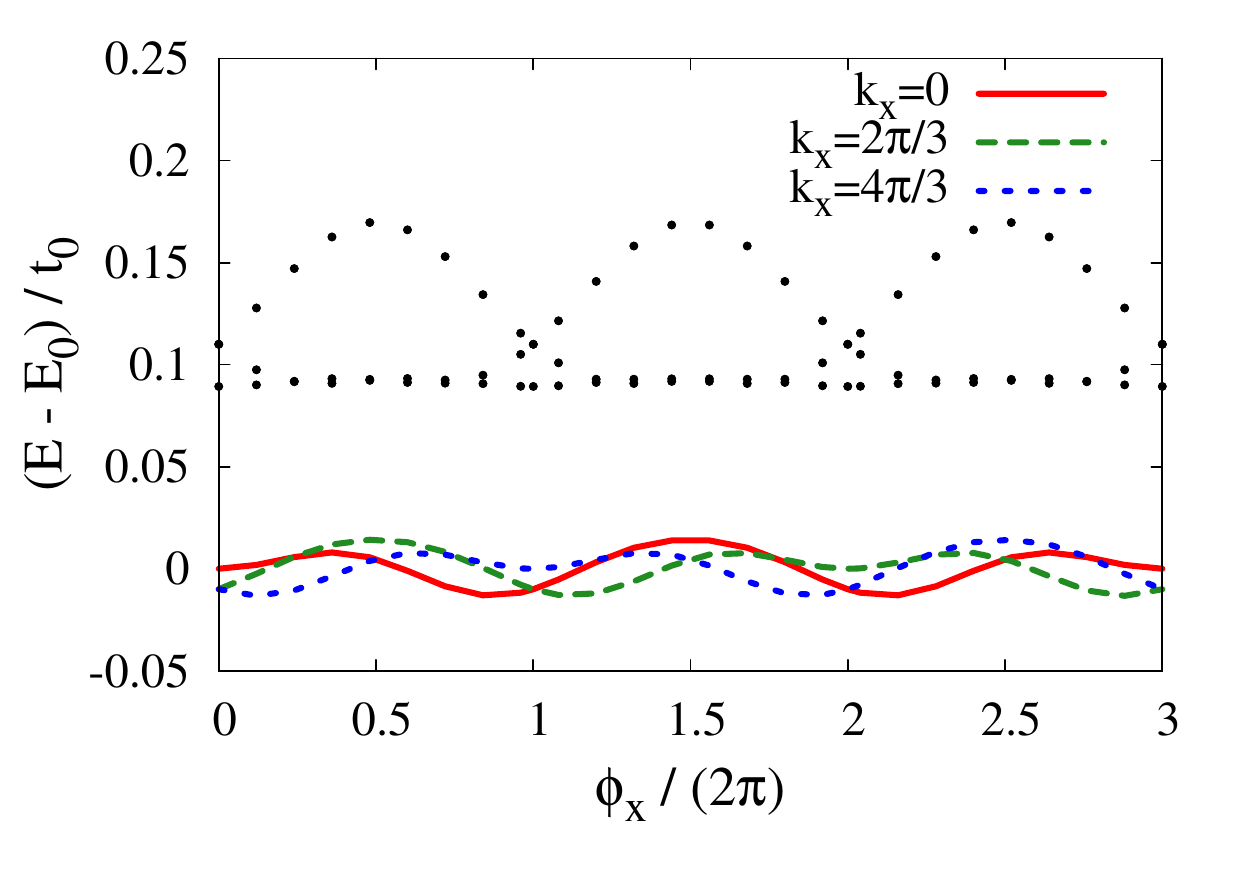}\label{fig:D15_U2_V1}}
\subfigure[$\;\Delta=1$, $U=2$, $V=1$]
{\includegraphics[width=0.47\columnwidth]
{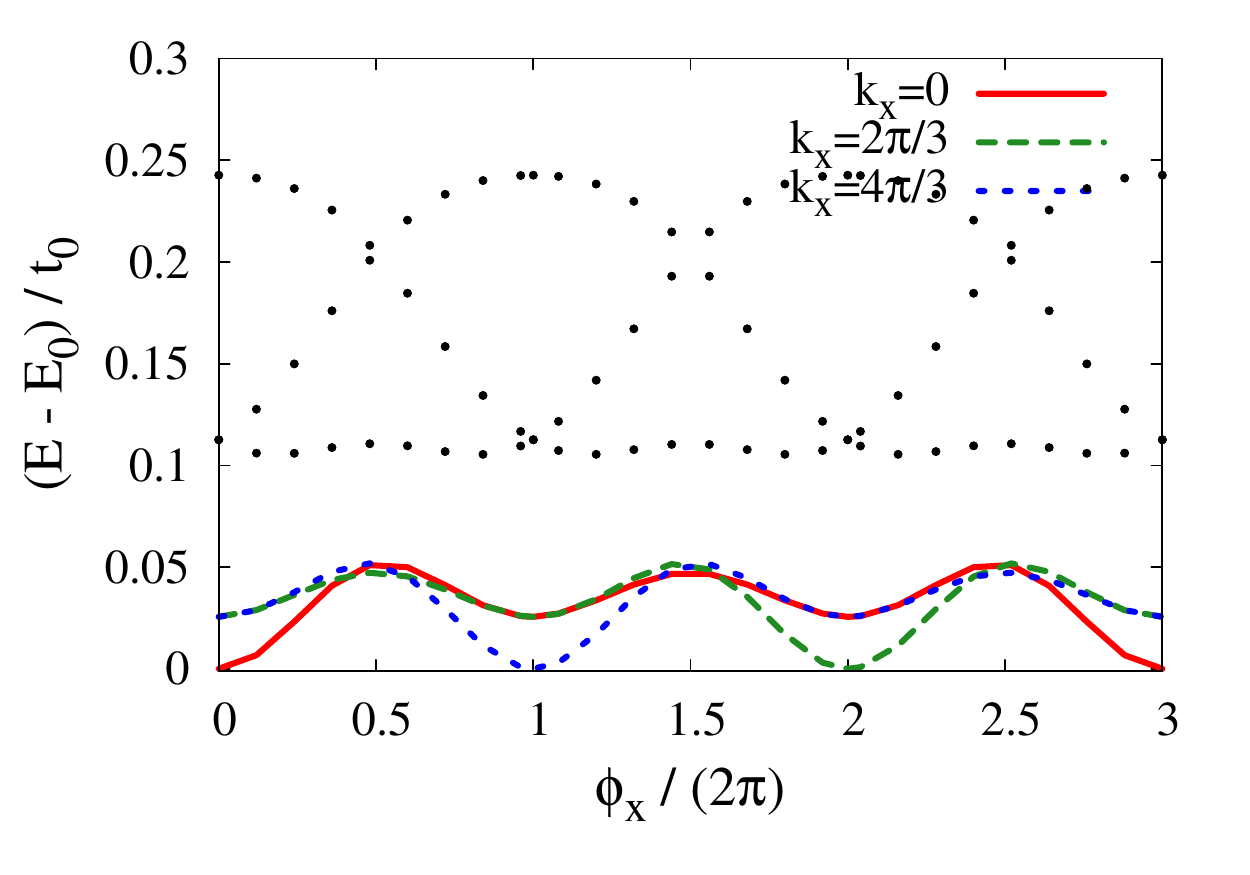}\label{fig:D1_U2_V1}}
\caption{\label{fig:ed_phi} (Color online) Lowest energies of the
  interacting $e_g$ system on the triangular lattice, depending on
  flux inserted for various parameter sets. (c) and (e) are given in
  the main text as Fig. 4(a,b). The ED results presented here were
  obtained for a $2\times 6 $ system with $t'/t=-0.46$.} 
\end{figure}

The
Hamiltonian is invariant under 
translations, hence total momentum constitutes a good quantum
number. Figure~\ref{fig:ed_k} shows the lowest two eigenvalues within
the sectors corresponding to the total 
momenta possible on a $6\times 1$ chain (the two sites in
$y$-direction define the two-site unit cell in real space), and one clearly sees three
low-energy states separated from the remaining spectrum by a gap. 
Adiabatic flux insertion is implemented by dividing the total
flux $\phi$ in smaller phases $\phi/N_x$ added to each hopping process, which is
equivalent to using twisted boundary conditions and preserves
translational invariance. This replaces the two hopping matrices whose
corresponding lattice vectors have a component along ${\bf a_1}$, i.e.,
$T_1$ and $T_3$ by $T_1\textrm{e}^{i\phi/N_x}$ and $T_3\textrm{e}^{i\phi/N_x}$. Figure 4 in the main text
shows the evolution of the ground state manifold as function of the
inserted flux. The system assumes an equivalent, yet not identical,
state after one unit $\phi=2\pi$ of flux is threaded through the system, meaning
that states in the ground state manifold have switched places. After 3
periods of flux insertion, where 3 corresponds to the inverse of the
filling fraction, $\nu^{-1}$, with $\nu=1/3$, we recover an identical
situation as zero flux. Figure~\ref{fig:ed_phi} here in the supplemental
information here shows the same for various other parameter sets,
indicating that the behavior is stable. 
This level crossing of the degenerate many
body ground state manifold reveals the existence of the FQH ground
state, provided that the spectral gap persists in the thermodynamic
limit, in order to unambiguously exclude a charge density wave.~\cite{regnault11} We cannot address
larger system sizes here, but some progress can hopefully be made in the future.

\bibliography{flat_bands_submission_final.bbl}

\begin{thebibliography}{44}
\bibitem [{not({\natexlab{b}})}]{note_hopp_JT}%
  \BibitemOpen
  \href@noop {} {} \bibinfo {note} {A distortion of the
  octahedra can induce hopping via the oxygen ligands along some directions, we
  verified that the total hoppings would have to increase by an unrealistic
  amount ($\approx 50\%$) before the band flattening is affected}\BibitemShut
  {NoStop}%
\end{thebibliography}

\end{document}